\documentclass[lettersize,journal]{IEEEtran}
\usepackage{amsmath,amsfonts}
\usepackage{algorithm}
\usepackage{algorithmic}
\usepackage{array}
\usepackage{booktabs}
\usepackage{multirow}
\usepackage[caption=false,font=normalsize,labelfont=sf,textfont=sf]{subfig}
\usepackage{textcomp}
\usepackage{stfloats}
\usepackage{url}
\usepackage{verbatim}
\usepackage{graphicx}
\usepackage{balance}

\begin{document}

\title{Fast Iterative Dual-Output-Aware LUT Mapping for Fracturable FPGA Architectures}

\author{Sheng Lu, Department of Electrical Engineering, The University of Texas at Arlington}

\maketitle

\begin{abstract}
Modern field-programmable gate arrays employ fracturable lookup tables (LUTs), each of which can implement either a single large-input function or two smaller functions within a shared physical LUT site. Conventional technology-mapping flows typically perform dual-output packing only after single-output LUT mapping, thereby preventing potential pairing opportunities from influencing cut selection. This paper presents an iterative dual-output-aware LUT-mapping framework integrated into the delay, area-flow, and exact-area optimization passes of the Berkeley ABC mapper. The proposed method alternates between single-output cut selection and bounded dual-output matching. Candidate pairs are efficiently generated using a sparse support-based index, validated against parameterized architectural, dependency, and output-specific timing constraints, and selected through a heuristic score-based matching procedure. The resulting partner information is then fed back into subsequent mapping rounds through a compatibility-aware cut-cost adjustment. To preserve timing accuracy, the physical union of the inputs is used only for architectural legality checking, while each output retains its own logical timing support. Experiments on the EPFL combinational benchmark suite under two representative architecture models demonstrate average reductions of 34.96\% and 23.39\% in the reported LUT-area metric relative to vanilla ABC. Compared with the best previously reported method, the proposed framework achieves a $15.8\times$ speedup while further reducing depth by approximately 5\% and area by 1\%. 
\end{abstract}


\section{Introduction}
\label{sec:introduction}

Look-up table (LUT) technology mapping transforms a technology-independent
Boolean network into an FPGA implementation~\cite{Cong_2011,SmartGrid_2016,Timpe_2023}. Because mapping determines the
number of logic resources and the depth of the resulting network, it directly
affects area, timing, power, and routability~\cite{Betz_1999,Ahmed_2000,Lamoureux_2006,Li_2018,Elgammal_2022,Ababei_2005}. Classical methods formulate the
problem with $K$-feasible cuts and optimize delay, area flow, or exact area
through iterative mapping rounds~\cite{Cong_1994,Mishchenko_2006,Brayton_2010}. These methods generally assume that each LUT
implements one output function.

Modern FPGAs increasingly use fracturable LUTs~\cite{Tan_2018,Murray_2020,Zhenghong_Jiang_2014}. A physical LUT site may
implement either one large function or two smaller compatible functions that
share input resources. Mapping two functions into one site can reduce logic
utilization and may also improve local interconnect usage. This opportunity has
motivated post-mapping packing, dual-output-aware cut selection, and general
multi-output mapping formulations~\cite{Jang_2009,Wang_2020,Calvino_2023,Hu_2008,Lee_2010}.

A common approach first obtains a high-quality single-output cover and then
merges compatible LUTs. Although computationally efficient, this strategy is
restricted by decisions made before packing. Two roots may have alternative
cuts that form a beneficial pair, but these cuts may never be selected when
each root is optimized independently. At the other extreme, explicit
multi-output cut enumeration exposes more alternatives but substantially
increases the search space and runtime. A practical mapper must therefore
balance mapping quality against candidate-generation and matching cost.
Existing general multi-output formulations provide a principled search model~\cite{Calvino_2023,Shi_2026},
but their larger cut spaces and more complex cost evaluation can limit their
benefit on large networks. This motivates a narrower integration in which the
well-established single-output mapper remains responsible for cut enumeration,
while a separate bounded stage identifies physical sharing opportunities and
returns only their estimated benefit to later recovery rounds.

Timing evaluation introduces a second difficulty. The outputs of a
fracturable LUT share a physical input interface, but they do not necessarily
have identical logical dependencies. Using the union of both supports as the
timing support of each output creates false timing arcs. A dual-output mapper
must distinguish physical input compatibility from output-specific logical
timing support.

This work couples conventional cut selection with bounded dual-output pairing. After each area-recovery round, the selected cuts form a sparse compatibility graph, where legal and timing-feasible pairs are scored and matched. The estimated site benefit of committed pairs is then fed back into subsequent cut-selection rounds, favoring cuts that are both individually competitive and jointly compatible. This approach exposes site-sharing opportunities before the final cover is fixed while preserving the scalability of priority-cut mapping. Although heuristic and sensitive to candidate bounds, the initial cover, and graph retention, it recovers useful dual-output opportunities without exhaustive multi-output cut enumeration.

The main contributions are as follows:
\begin{itemize}
    \item An iterative mapping framework that feeds dual-output compatibility
    and physical-site benefit into subsequent single-output cut-selection
    rounds.
    \item A parameterized legality model that separates per-output support,
    physical input-union capacity, optional sharing constraints, root
    independence, and output-specific timing support.
    \item A bounded shared-input candidate search and sparse matching procedure
    that controls runtime while retaining high-value pairing opportunities.
\end{itemize}

\section{Background}
\label{sec:background}

\subsection{Fracturable and Dual-Output LUTs}

\begin{figure}[!t]
    \centering
    \includegraphics[width=1.0\linewidth]{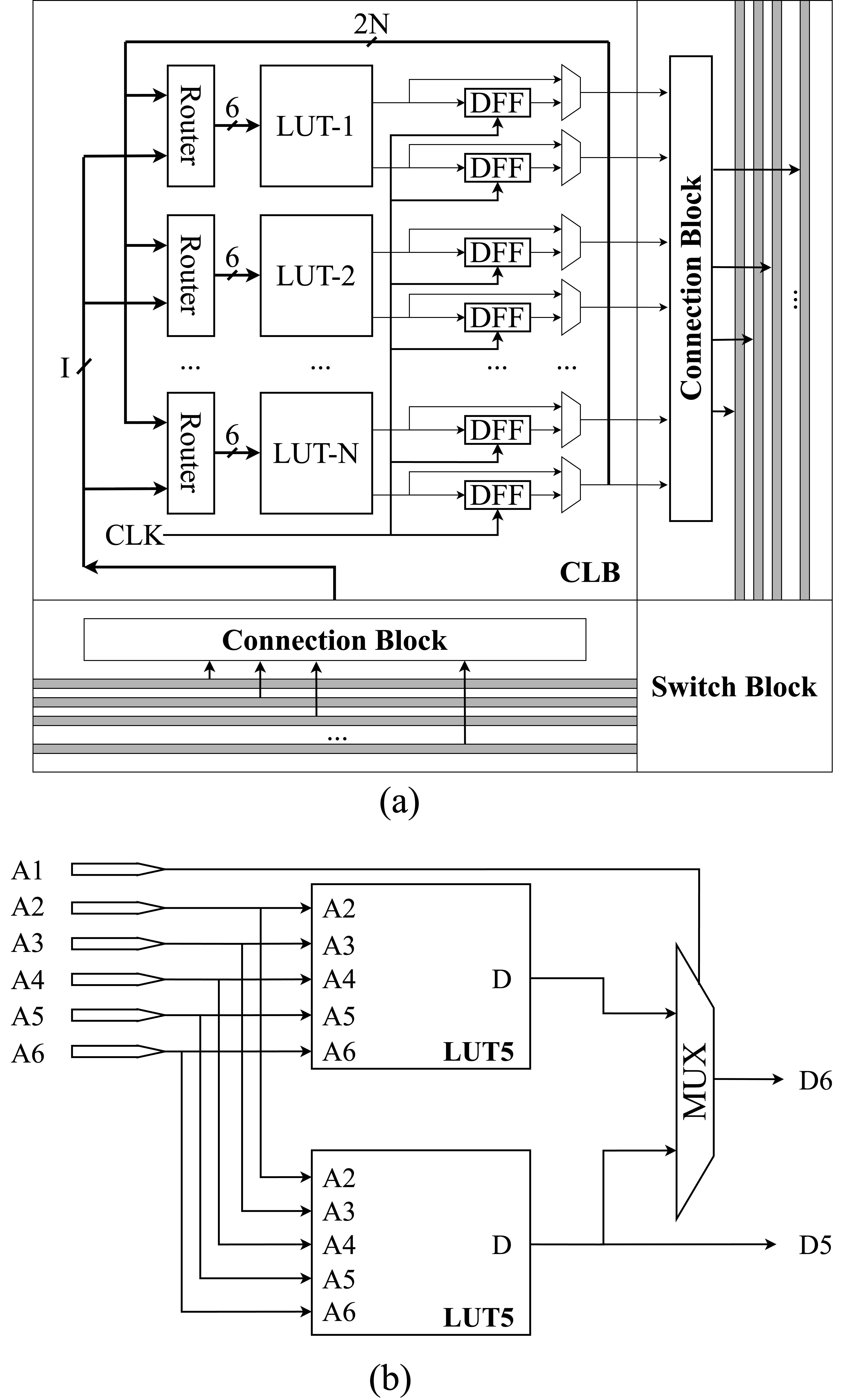}
    \caption{(a) FPGA configurable logic block and (b) representative fracturable
    LUT6 dual-output structure.}
    \label{fig:fig1}
\end{figure}

Modern FPGAs are commonly organized as arrays of configurable logic blocks
(CLBs) connected by programmable routing resources. As shown in
Fig.~\ref{fig:fig1}(a), a CLB contains multiple LUT sites, local input routers,
flip-flops, output multiplexers, and interfaces to the routing fabric through
connection and switch blocks. The external routing channels deliver signals to
the CLB input pins, denoted by $I$, and the local routers select up to six
inputs for each LUT. The parameter $N$ denotes the number of LUT sites in the
CLB, and since each LUT site can provide two output paths, the CLB can expose
up to $2N$ local outputs. Each LUT output can either bypass or pass through a
D flip-flop (DFF) driven by $\mathrm{CLK}$ before entering the
connection block. Therefore, LUT mapping affects not only the number of used LUT sites, but also local
output usage, sequential resource usage, and routing pressure inside the FPGA
fabric.

A key feature of commercial FPGA LUTs is fracturability. A fracturable LUT can
operate either as one large single-output LUT or as two smaller LUTs sharing
part of their input set. Fig.~\ref{fig:fig1}(b) shows a representative
dual-output LUT6 structure. Inputs $A_1,\ldots,A_6$ are available to the
physical LUT. The upper output $D_6$ can implement a six-input function, while
the lower output $D_5$ can expose one five-input subfunction. In dual-output
mode, the structure can be viewed as two LUT5 blocks that share inputs
$A_2,\ldots,A_6$. The signal $A_1$ is used to select between the two LUT5
outputs for the $D_6$ output through the internal multiplexer. Meanwhile, one
of the LUT5 outputs can be directly observed at $D_5$.

Commercial FPGA families commonly exploit this type of structure~\cite{Tan_2018,Timpe_2023}. For example,
a LUT6 can implement either one six-input function or two five-input functions
when the two functions share the required input subset. This is the basic
dual-output mode used in many LUT6-based architectures. Some newer FPGA
architectures provide more flexible fracturable LUT resources~\cite{Lu_2024,Lu_2025_RFET,Lu_2025_TCSO,Gaillardon_2015,Amaru_2013}, allowing two
outputs with larger compatible input supports, but the same principle remains:
multiple logic functions may share one physical LUT only when their individual
supports and their union support satisfy the architectural constraints.

We describe a target architecture by
\begin{equation}
\mathcal{A}=(K_{\mathrm{single}},K_{\mathrm{out}},K_{\mathrm{union}},
S_{\min},S_{\max}),
\label{eq:architecture_model}
\end{equation}
where $K_{\mathrm{single}}$ is the single-output input limit,
$K_{\mathrm{out}}$ is the support limit of each function in dual-output mode,
$K_{\mathrm{union}}$ is the physical input-union limit, and $S_{\min}$ and
$S_{\max}$ optionally bound the number of shared inputs.  When an architecture
does not impose an explicit sharing constraint, we use $S_{\min}=0$ and
$S_{\max}=K_{\mathrm{out}}$.

For candidate cuts $C_u$ and $C_v$, let
$I_u=\operatorname{supp}(C_u)$ and $I_v=\operatorname{supp}(C_v)$.  The pair
is architecture legal only if
\begin{equation}
\begin{aligned}
|I_u| &\leq K_{\mathrm{out}}, &
|I_v| &\leq K_{\mathrm{out}},\\
|I_u \cup I_v| &\leq K_{\mathrm{union}}, &
S_{\min} &\leq |I_u\cap I_v|\leq S_{\max}.
\end{aligned}
\label{eq:arch_pair_limits}
\end{equation}
The roots must also be mutually independent. If $r_u$ and $r_v$ are the roots
of $C_u$ and $C_v$, a legal pair must satisfy
\begin{equation}
r_u \notin I_v,\qquad r_v \notin I_u.
\label{eq:arch_root_independence}
\end{equation}
These parameterized constraints allow the same mapping formulation to target
different fracturable FPGA architectures.

\subsection{$K$-Feasible Cut Enumeration for Single-Output LUT}

Traditional LUT technology mapping in ABC is based on $K$-feasible cut
enumeration~\cite{Cong_1994,Mishchenko_2006,Brayton_2010}. Given a combinational network represented as a directed acyclic
graph $G=(V,E)$, each internal node $v \in V$ is treated as a possible LUT
root. A cut $C_v$ of node $v$ is a set of nodes that separates $v$ from the
primary inputs. Equivalently, every path from any primary input to $v$ must
pass through at least one node in $C_v$. If the cut contains no more than $K$
nodes, it is called a $K$-feasible cut and can be implemented by one
$K$-input LUT:
\begin{equation}
|C_v| \leq K.
\label{eq:k_feasible_cut}
\end{equation}

\begin{figure}[!t]
    \centering
    \includegraphics[width=0.7\linewidth]{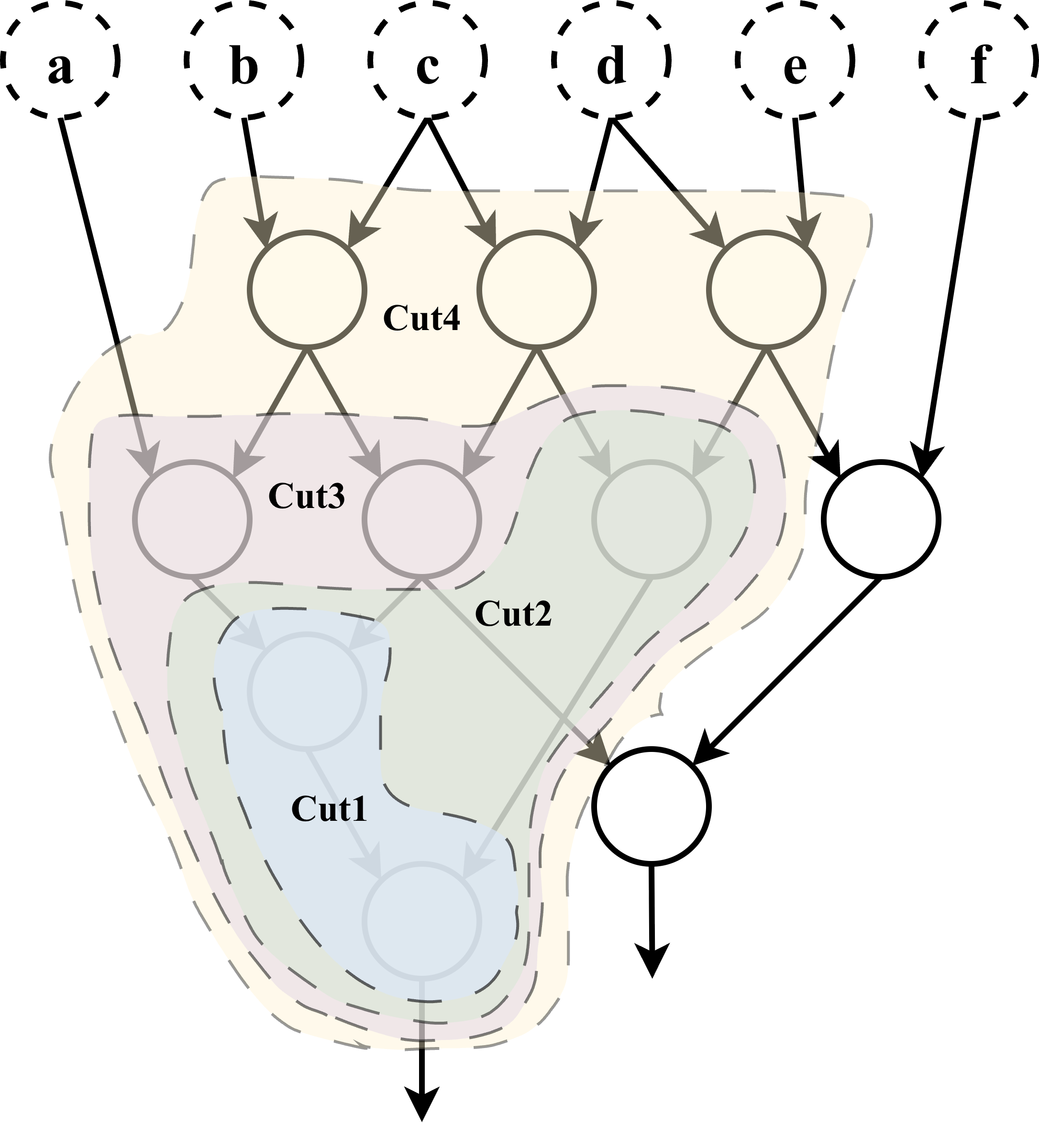}
    \caption{Illustration of multiple $K$-feasible cuts rooted at the same
    logic node in conventional single-output LUT mapping. The shaded regions
    denote cut-induced logic cones, while each cut is formally defined by its
    boundary leaf set.}
    \label{fig:fig2}
\end{figure}

Fig.~\ref{fig:fig2} illustrates several feasible cuts rooted at the same
logic node. The dashed nodes at the top, labeled $a$--$f$, represent primary
inputs or previously mapped signals. The solid nodes represent internal logic
nodes in the transitive fanin cone of the target root. Strictly, a cut is the
boundary node set that separates the root from the primary inputs; the shaded
regions in the figure illustrate the logic cones induced by different
candidate cuts. For example, $\mathrm{Cut}_1$ represents a small local cone
close to the root, while $\mathrm{Cut}_4$ represents a larger cone that
reaches farther toward the primary inputs. These alternative cut-induced cones
expose different possible LUT implementations for the same root node.

During cut enumeration, cuts are generated bottom-up following the topological
order of the network. For a two-input logic node $v$ with fanins $u_0$ and
$u_1$, candidate cuts of $v$ are produced by merging cuts from its fanins,
retaining only $K$-feasible merged cuts; the trivial cut may also be considered:
\begin{equation}
C_v=C_{u_0}\cup C_{u_1},\quad
|C_{u_0}\cup C_{u_1}|\leq K,\quad
C_v=\{v\}.
\label{eq:cut_construction}
\end{equation}
This procedure produces multiple candidate cuts for each node, corresponding
to different possible LUT boundaries.

Because the number of feasible cuts can grow rapidly, practical mappers keep
only a bounded set of priority cuts for each node. Let $\mathcal{C}(v)$ denote
the set of retained cuts for node $v$. ABC limits its size by a parameter
$n_{\mathrm{cuts}}$, such that $|\mathcal{C}(v)| \leq n_{\mathrm{cuts}}$.
The retained cuts are ranked according to mapping objectives such as delay,
area flow, exact area, edge count, or tie-breaking cost. Therefore, cut
enumeration is not only a feasibility procedure, but also the foundation for
optimization: different retained cuts expose different trade-offs between
logic depth and resource usage.

For each feasible cut, the mapper computes timing and area-related metrics.
The delay of a cut is determined by the latest arrival time among its leaves
plus the delay of the LUT implementing the cut:
\begin{equation}
\mathrm{arr}(v,C_v)
=
\max_{u \in C_v} \mathrm{arr}(u)
+
d_{\mathrm{LUT}}(C_v).
\label{eq:cut_arrival}
\end{equation}
A delay-oriented mapping round selects cuts that minimize the arrival time of
each node. After the initial delay mapping, ABC performs area-flow and
exact-area recovery rounds. Area flow estimates the fractional contribution of
a cut by distributing the cost of shared fanin logic across its fanouts, while
exact-area recovery evaluates the incremental area more directly by
temporarily dereferencing and referencing mapped cones.

Finally, for conventional single-output mapping, one best cut is selected for
each mapped root. The selected cut defines one single-output LUT whose input
pins are the leaves of the cut and whose output is the root node. Thus, the
standard ABC mapping flow constructs a network of single-output LUTs,
$v \leftarrow \mathrm{LUT}(C_v)$, where each LUT implements one Boolean function and satisfies the
$K$-input constraint in Eq.~\eqref{eq:k_feasible_cut}. This single-output cut
enumeration and selection process is the baseline on which the proposed
dual-output mapping method is built.

\section{Proposed Method}
\label{sec:method}

\subsection{Overview of Cut-based Mapping Flow}
\label{subsec:overview_flow}

\begin{figure}[!t]
    \centering
    \includegraphics[width=1.0\linewidth]{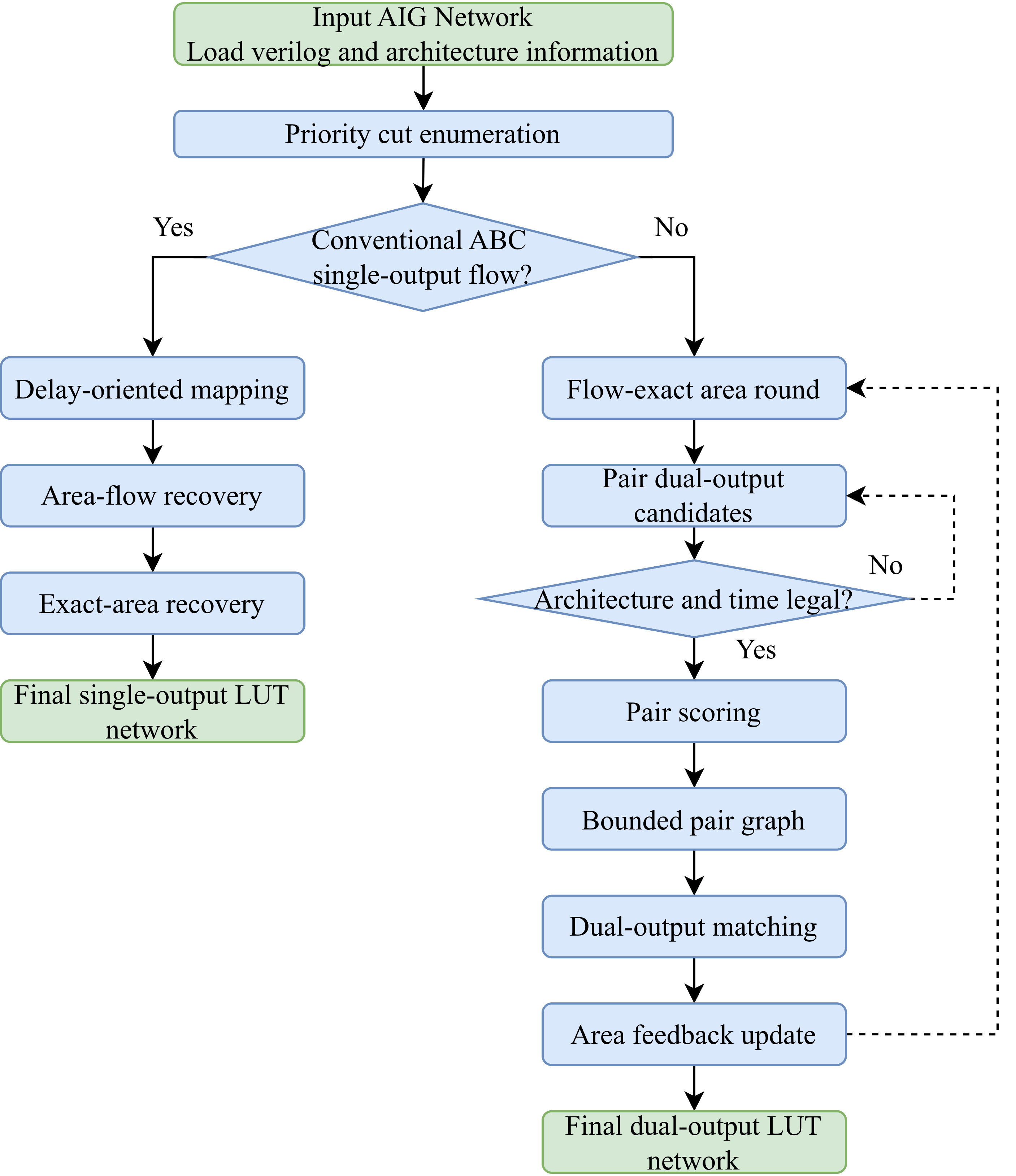}
    \caption{Overview of the proposed iterative dual-output-aware mapping
    flow compared with the conventional ABC single-output mapping flow.}
    \label{fig:fig3}
\end{figure}

Fig.~\ref{fig:fig3} summarizes the proposed dual-output-aware mapping flow and
its relationship to the conventional ABC single-output mapper. The input
network is first converted into an and-inverter graph (AIG), and
architecture information is loaded to specify the LUT capacity and
dual-output legality constraints. The mapper then performs priority cut enumeration, which provides the candidate
single-output cuts used by both the conventional and proposed flows.

In the conventional ABC flow, the mapper proceeds linearly after cut
enumeration. It first performs delay-oriented mapping to obtain a timing-driven
initial cover, then applies area-flow recovery and exact-area recovery to
reduce LUT usage under timing constraints. The final result is a
single-output LUT network in which each mapped LUT implements one output
function.

The proposed flow keeps the ABC cut enumeration and mapping rounds, but embeds
dual-output matching into the iterative mapping process. After each flow or
exact-area round, the currently referenced mapped roots are treated as
single-output groups. The algorithm then constructs candidate dual-output
pairs from these groups. Candidate generation is guided by shared cut inputs:
LUTs that use common input nodes are placed into shared-input buckets, and
only a bounded number of related candidates are evaluated. This avoids
exhaustive all-pair enumeration while still focusing the search on pairs that
are likely to share a physical fracturable LUT.

Each generated pair is then checked against the architecture and timing
constraints. A pair is rejected if either output cut exceeds the dual-output
input limit, if the union of the two input sets exceeds the physical LUT input
limit, if one root depends on the other root, or if the output-specific timing
constraint is violated. Illegal pairs are discarded, and the algorithm returns
to candidate generation for additional candidates. Legal pairs are scored
according to their expected area saving, shared-input quality, timing impact,
and structural complexity.

The legal and scored pairs form a sparse bounded pair graph. The mapper first
commits mutual-best pairs, where both endpoints select each other as their
best partner. It then greedily commits the remaining non-conflicting pairs in
descending score order, ensuring that each mapped root participates in at most
one dual-output group. After matching, the committed partner relations are
stored and used as feedback in the next mapping round. Specifically, cuts that
remain compatible with previously selected dual-output partners receive a
reduced area cost, encouraging subsequent ABC mapping rounds to preserve or
improve useful dual-output opportunities.

This feedback loop couples single-output cut selection with dual-output LUT
packing. Therefore, the proposed method is not merely a post-processing merge
of an already mapped single-output network. Instead, dual-output feasibility
and area benefit are iteratively propagated back into the cut-selection
process. The final output is a dual-output LUT network in which matched pairs
share physical fracturable LUT resources, while unmatched roots remain as
ordinary single-output LUTs.

\subsection{Bounded Candidate Generation}
\label{subsec:pair_elimination}

\begin{figure}[!t]
    \centering
    \includegraphics[width=1\linewidth]{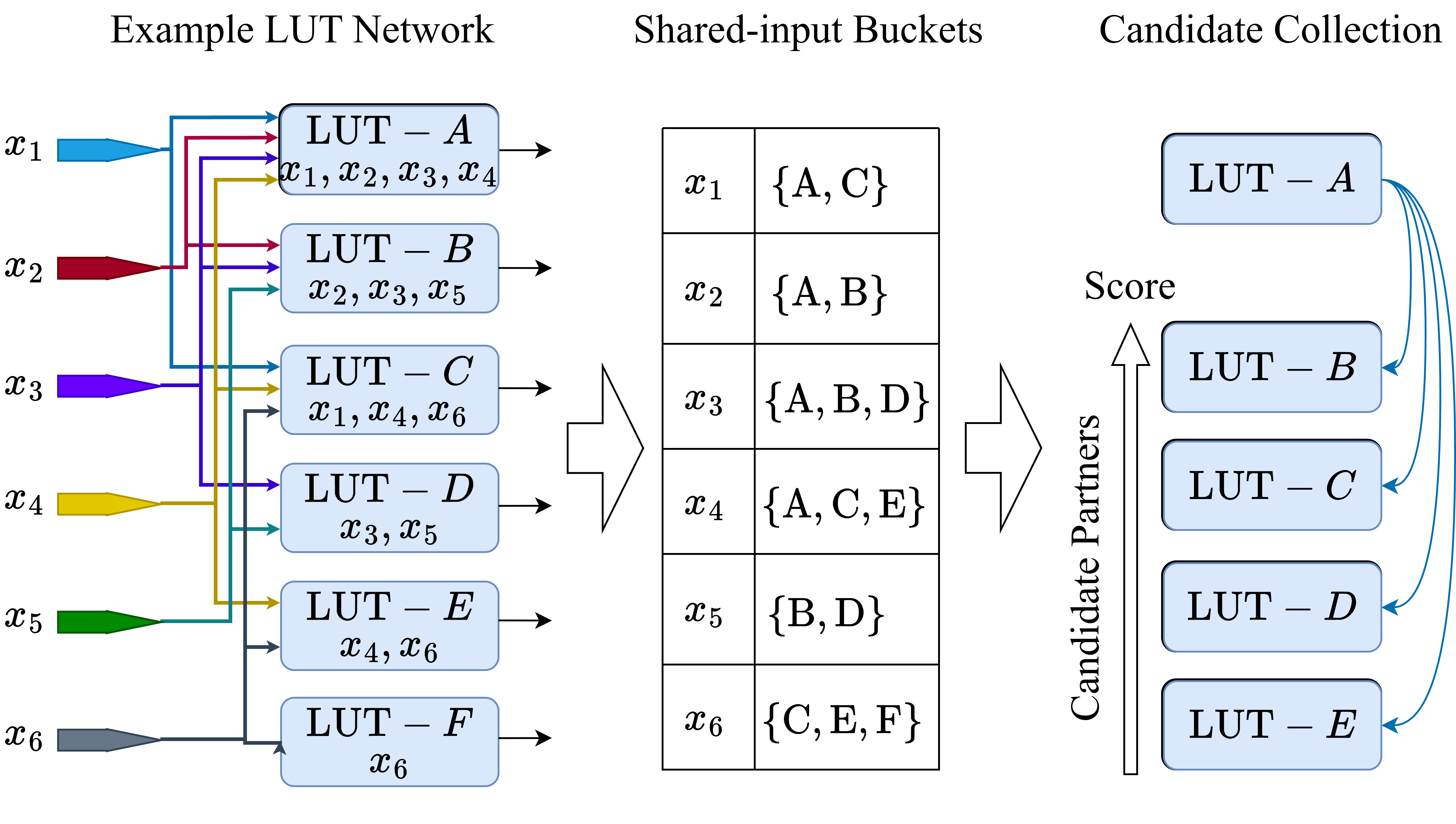}
    \caption{Shared-input bucket based bounded pruning for dual-output LUT
    candidate generation. LUTs are indexed by their input supports, and only
    LUTs sharing inputs with the target LUT are collected as candidate
    partners before legality checking and scoring.}
    \label{fig:fig4}
\end{figure}

A direct way to identify dual-output LUT pairs is to test every pair of
currently referenced LUTs. However, this all-pair strategy is expensive. If
there are $m$ referenced LUTs in the mapped network, exhaustive pair checking
requires $\mathcal{O}(m^2)$ candidate evaluations. Most of these pairs are unlikely to be useful because
pairs with common inputs tend to have a smaller physical input union.
Therefore, the proposed mapper uses heuristic bounded pruning based on
shared-input buckets.

Fig.~\ref{fig:fig4} illustrates the basic idea. Each currently referenced LUT
is first represented by its input support. For example, the target
$\mathrm{LUT}\text{-}A$ uses $I_A = \{x_1,x_2,x_3,x_4\}$.
Instead of comparing $\mathrm{LUT}\text{-}A$ with all other LUTs, the mapper
builds an inverse index from each input signal to the LUTs that use this
input:
\begin{equation}
\mathcal{B}(x_i)
=
\{\,L \mid x_i \in I_L\,\},
\label{eq:shared_input_bucket}
\end{equation}
where $\mathcal{B}(x_i)$ is the shared-input bucket of input $x_i$ and $I_L$
is the input support of LUT $L$.

For the example in Fig.~\ref{fig:fig4}, the buckets associated with the
inputs of $\mathrm{LUT}\text{-}A$ are
\begin{equation}
\begin{aligned}
\mathcal{B}(x_1)&=\{A,C\}, &
\mathcal{B}(x_2)&=\{A,B\},\\
\mathcal{B}(x_3)&=\{A,B,D\}, &
\mathcal{B}(x_4)&=\{A,C,E\}.
\end{aligned}
\label{eq:example_buckets}
\end{equation}
The raw candidate partner set of $\mathrm{LUT}\text{-}A$ is then obtained by
taking the union of these buckets and removing $A$ itself:
\begin{equation}
\mathcal{P}_A
=
\left(
\bigcup_{x_i \in I_A} \mathcal{B}(x_i)
\right)
\setminus \{A\}
=\{B,C,D,E\}.
\label{eq:candidate_union}
\end{equation}

This bucket-based construction omits pairs that do not share any input
with the target LUT. For instance, $\mathrm{LUT}\text{-}F$ is not considered
as a candidate partner for $\mathrm{LUT}\text{-}A$ because it only uses
$x_6$, which is not in $I_A$. In this way, the mapper avoids evaluating
pairs before applying the more expensive architecture and timing legality
checks.  Shared input, however, is not a necessary condition for legality when
$S_{\min}=0$: two disjoint supports may still satisfy
$|I_u\cup I_v|\leq K_{\mathrm{union}}$.  Consequently, the current generator
is heuristic and does not enumerate every architecture-legal pair.  In this
paper, bounded candidate pruning denotes this quality--runtime trade-off; it
is not an exact elimination rule and provides no completeness guarantee.

To control runtime, two search limits are introduced. The first limit,
$n_{\mathrm{bucket}}$, bounds the number of entries scanned from each
shared-input bucket. The second limit, $n_{\mathrm{cand}}$, bounds the number
of distinct candidate partners evaluated for each target LUT:
\begin{equation}
|\mathrm{scan}(\mathcal{B}(x_i))|\leq n_{\mathrm{bucket}},\qquad
|\mathcal{P}_A^{\mathrm{eval}}|\leq n_{\mathrm{cand}}.
\label{eq:candidate_search_limits}
\end{equation}
These two parameters correspond to the implementation options
$n_{\mathrm{BucketScanMax}}$ and $n_{\mathrm{CandidatesMax}}$, respectively.

Algorithm~\ref{alg:pair_generation} gives the corresponding implementation.
For bounded support size $K$, bucket construction costs $\mathcal{O}(mK)$ and
candidate scanning costs $\mathcal{O}(mK n_{\mathrm{bucket}})$, capped by
$\mathcal{O}(m n_{\mathrm{cand}})$ distinct evaluated candidates~\cite{Beamer_2013}.

\begin{algorithm}[!t]
\caption{Bounded Shared-Input Candidate Generation}
\label{alg:pair_generation}
\begin{algorithmic}[1]
\STATE \textbf{Input:} referenced LUT roots $R$, supports $I_u$, limits
$n_{\mathrm{bucket}}$, $n_{\mathrm{cand}}$
\STATE \textbf{Output:} candidate partner sets $\mathcal{P}_u$
\FOR{each input $x$}
    \STATE $\mathcal{B}(x)\leftarrow\emptyset$
\ENDFOR
\FOR{each root $u\in R$}
    \FOR{each input $x\in I_u$}
        \STATE Insert $u$ into $\mathcal{B}(x)$
    \ENDFOR
\ENDFOR
\FOR{each root $u\in R$}
    \STATE $\mathcal{P}_u\leftarrow\emptyset$
    \FOR{each input $x\in I_u$}
        \STATE Scan at most $n_{\mathrm{bucket}}$ entries of $\mathcal{B}(x)$
        \FOR{each scanned root $v$}
            \IF{$v\neq u$}
                \STATE Insert $v$ into $\mathcal{P}_u$ if it is new
            \ENDIF
            \IF{$|\mathcal{P}_u|=n_{\mathrm{cand}}$}
                \STATE Stop collecting candidates for $u$
            \ENDIF
        \ENDFOR
    \ENDFOR
\ENDFOR
\RETURN $\{\mathcal{P}_u\mid u\in R\}$
\end{algorithmic}
\end{algorithm}

After candidate collection, the remaining candidates are not immediately
merged. Each candidate pair is first checked against the dual-output legality
constraints, including individual cut-size limits, union input limits,
root-independence constraints, and output-specific timing validity. Legal pairs
are then scored. A simplified score for a target LUT $u$ and a candidate LUT
$v$, consistent with the detailed scoring model in Section~\ref{subsec:pair_scoring}, can be written as
\begin{equation}
S(u,v)
=
\alpha \Delta A(u,v)
+
\beta |I_u \cap I_v|
-
\eta |I_u \cup I_v|
-
\gamma \Delta D(u,v),
\label{eq:pair_score}
\end{equation}
where $\Delta A(u,v)$ is the estimated physical LUT area saving,
$|I_u \cap I_v|$ rewards shared inputs, $|I_u \cup I_v|$ penalizes larger
merged input support, and $\Delta D(u,v)$ captures timing degradation or
timing risk. The coefficients $\alpha$, $\beta$, $\eta$, and $\gamma$
control the relative importance of these terms.

Finally, only high-scoring legal pairs are retained in the bounded pair graph.
For each LUT, the mapper keeps at most $n_{\mathrm{pair}}$ candidate edges,
$|\mathcal{E}_u| \leq n_{\mathrm{pair}}$, where $\mathcal{E}_u$ denotes the retained legal pair edges incident to LUT
$u$. This final pruning step further reduces matching complexity while
preserving the most promising dual-output opportunities.

\subsection{Pair-Compatibility Feedback to Cut Selection}
\label{subsec:pair_feedback}

Let $\mathcal{M}^{(t-1)}$ be the nonconflicting pair set produced after the
previous recovery round, and let $p^{(t-1)}(u)$ denote the partner of root $u$
in that set.  For a candidate cut $C_u$ considered in round $t$, define
\begin{equation}
\chi(C_u,p^{(t-1)}(u))=
\begin{cases}
1, & \text{if $C_u$ is legal with cut of $p^{(t-1)}(u)$},\\
0, & \text{otherwise.}
\end{cases}
\label{eq:feedback_compatibility}
\end{equation}
The estimated benefit of retaining that compatibility and the resulting
feedback-adjusted area-flow and exact-area ranking costs are
\begin{equation}
\begin{aligned}
B(C_u,C_p)&=\max\left(0,A(C_u)+A(C_p)-A_{\mathrm{dual}}(C_u,C_p)\right),\\
\operatorname{AF}_{\mathrm{fb}}(C_u)&=
\operatorname{AF}_{\mathrm{ABC}}(C_u)
-\chi(C_u,p^{(t-1)}(u))B(C_u,C_p),\\
\operatorname{EA}_{\mathrm{fb}}(C_u)&=
\operatorname{EA}_{\mathrm{ABC}}(C_u)
-\chi(C_u,p^{(t-1)}(u))B(C_u,C_p).
\end{aligned}
\label{eq:feedback_costs}
\end{equation}
The original ABC delay constraint is applied before this adjustment, so
feedback changes only the ranking of delay-feasible cuts and cannot make an
illegal or delay-infeasible cut selectable.

Algorithm~\ref{alg:iterative_mapping} summarizes the outer loop.  The current
implementation terminates after the predefined ABC area-flow and exact-area
recovery-round budgets are exhausted.  All pairs are checked again against the
final cuts before they are committed.

\begin{algorithm}[!t]
\caption{Iterative Dual-Output-Aware Mapping}
\label{alg:iterative_mapping}
\begin{algorithmic}[1]
\STATE \textbf{Input:} AIG $G$, architecture $\mathcal{A}$, recovery-round
budget $T$
\STATE \textbf{Output:} mapped LUT network and final pair set $\mathcal{M}$
\STATE Enumerate priority cuts and run the original delay-oriented mapping
\STATE $\mathcal{M}\leftarrow\emptyset$
\FOR{$t=1$ to $T$ over area-flow and exact-area rounds}
    \STATE Filter cuts using the original ABC delay constraint
    \STATE Rank feasible cuts using Eq.~\eqref{eq:feedback_costs}, as applicable
    \STATE Select one cut for each referenced root
    \STATE Generate bounded candidates and discard illegal pairs
    \STATE Score legal edges and compute a nonconflicting set $\mathcal{M}'$
    \STATE $\mathcal{M}\leftarrow\mathcal{M}'$
\ENDFOR
\STATE Recheck $\mathcal{M}$ against the final selected cuts
\STATE Commit valid pairs; retain every other root as a single-output LUT
\RETURN mapped network and $\mathcal{M}$
\end{algorithmic}
\end{algorithm}

\subsection{Output-Specific Timing After Dual-Output Pairing}
\label{subsec:output_specific_timing}

\begin{figure}[!t]
    \centering
    \includegraphics[width=0.75\linewidth]{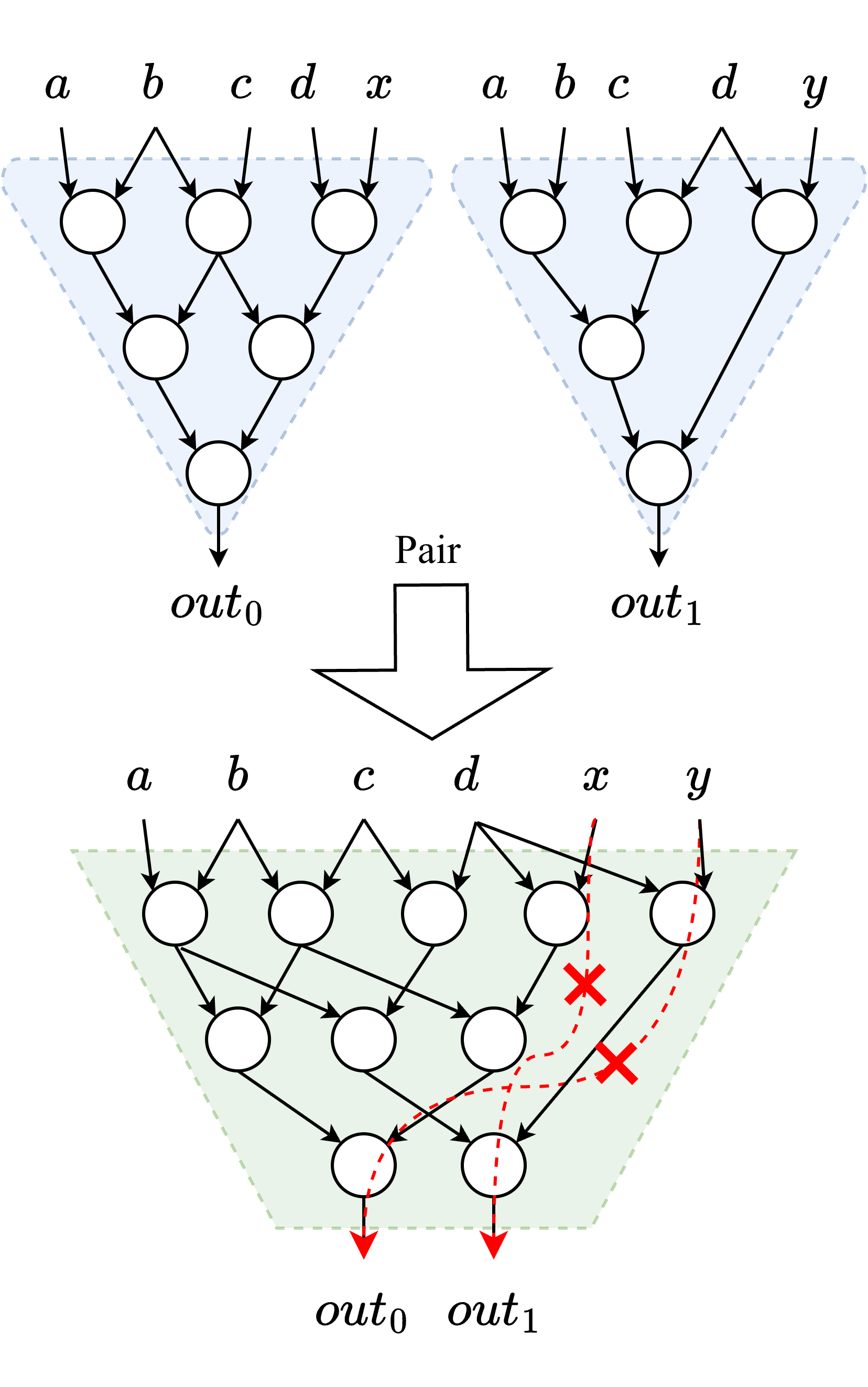}
    \caption{Output-specific timing interpretation after dual-output LUT
    pairing. The merged physical LUT receives the union of the two input sets,
    but each output is timed only through its own original cut. Red dashed
    paths indicate artificial timing paths that should not be introduced.}
    \label{fig:fig5}
\end{figure}

Dual-output LUT pairing changes the physical implementation of two mapped
cuts, but it should not change the logical dependency of each output. This is
an important distinction for timing analysis. Fig.~\ref{fig:fig5} illustrates
the issue. Before pairing, the two single-output LUT cones are independent.
The first output, $out_0$, depends on its own cut input set, while the second
output, $out_1$, depends on another cut input set. After pairing, the two cuts
are placed into one shared physical LUT structure whose input pins correspond
to the union of the two input sets. However, this physical union does not mean
that every union input becomes a timing predecessor of both outputs.

Let the two mapped cuts be $C_u$ and $C_v$, rooted at outputs $r_u$ and
$r_v$, respectively. Their input supports and the merged physical input set are
\begin{equation}
I_u=\mathrm{supp}(C_u),\qquad
I_v=\mathrm{supp}(C_v),\qquad
I_{uv}=I_u\cup I_v.
\label{eq:timing_supports}
\end{equation}
The set $I_{uv}$ is used only to check whether the pair fits into the target
fracturable LUT architecture. It should not be interpreted as the timing
support of both outputs.

For correct timing evaluation, each output must be timed through its own
original cut. Therefore, the arrival times are computed as
\begin{equation}
\begin{aligned}
\mathrm{arr}(r_u)&=\max_{p\in I_u}\mathrm{arr}(p)+d_u(C_u),\\
\mathrm{arr}(r_v)&=\max_{p\in I_v}\mathrm{arr}(p)+d_v(C_v).
\end{aligned}
\label{eq:paired_arrivals}
\end{equation}
Here, $d_u(C_u)$ and $d_v(C_v)$ denote the output-specific LUT delays for the
two outputs. These delays may be identical in a simplified LUT model, but they
are written separately to allow architecture-dependent output delays.

A common incorrect interpretation is to time both outputs using the full union
input set $I_{uv}$. This would imply
\begin{equation}
\begin{aligned}
\mathrm{arr}(r_u)&=\max_{p\in I_{uv}}\mathrm{arr}(p)+d_u(C_u),\\
\mathrm{arr}(r_v)&=\max_{p\in I_{uv}}\mathrm{arr}(p)+d_v(C_v).
\end{aligned}
\label{eq:incorrect_arrivals}
\end{equation}
However, this creates artificial timing paths. For example, an input that
belongs only to $I_v$ would be treated as a predecessor of $r_u$, and an input
that belongs only to $I_u$ would be treated as a predecessor of $r_v$. These
paths do not exist in the original Boolean functions and are marked as invalid
in Fig.~\ref{fig:fig5}.

Therefore, the proposed mapper separates physical compatibility from timing
dependency. The union set $I_u \cup I_v$ is used for architecture legality,
while timing is evaluated using the output-specific supports $I_u$ and $I_v$.
The pair is timing legal only if both outputs satisfy their required times:
\begin{equation}
\mathrm{arr}(r_u)\leq\mathrm{req}(r_u),\qquad
\mathrm{arr}(r_v)\leq\mathrm{req}(r_v).
\label{eq:required_times}
\end{equation}
This preserves the original logical timing dependencies under the evaluated
delay model while allowing the mapper to exploit the estimated site benefit
of sharing one dual-output LUT.

Algorithm~\ref{alg:legality_check} summarizes the complete legality test. It
combines the architectural constraints in
Eqs.~\eqref{eq:arch_pair_limits}--\eqref{eq:arch_root_independence} with the
output-specific timing conditions in Eqs.~\eqref{eq:paired_arrivals} and
\eqref{eq:required_times}.
For bounded LUT size, the test costs $\mathcal{O}(K)$ per candidate pair.

\begin{algorithm}[!t]
\caption{Dual-Output Legality and Timing Check}
\label{alg:legality_check}
\begin{algorithmic}[1]
\STATE \textbf{Input:} pair $(u,v)$, cuts $C_u,C_v$, roots $r_u,r_v$,
required times $\mathrm{req}(r_u),\mathrm{req}(r_v)$
\STATE \textbf{Output:} legal or illegal
\STATE $I_u\leftarrow\mathrm{supp}(C_u)$, $I_v\leftarrow\mathrm{supp}(C_v)$
\IF{$|I_u|>K_{\mathrm{out}}$ or $|I_v|>K_{\mathrm{out}}$}
    \RETURN illegal
\ENDIF
\IF{$|I_u\cup I_v|>K_{\mathrm{union}}$}
    \RETURN illegal
\ENDIF
\IF{$|I_u\cap I_v|<S_{\min}$ or $|I_u\cap I_v|>S_{\max}$}
    \RETURN illegal
\ENDIF
\IF{$r_u\in I_v$ or $r_v\in I_u$}
    \RETURN illegal
\ENDIF
\STATE Compute $\mathrm{arr}(r_u)$ and $\mathrm{arr}(r_v)$ with output-specific supports
\IF{$\mathrm{arr}(r_u)>\mathrm{req}(r_u)$ or $\mathrm{arr}(r_v)>\mathrm{req}(r_v)$}
    \RETURN illegal
\ENDIF
\RETURN legal
\end{algorithmic}
\end{algorithm}

\subsection{Dual-Output Pair Scoring}
\label{subsec:pair_scoring}

After bounded candidate generation and architecture legality checking, the mapper obtains
a set of legal dual-output candidate pairs. These pairs are not immediately
committed, because one LUT can participate in at most one dual-output group.
Therefore, the mapper assigns a score to each legal pair and constructs a
bounded pair graph for the subsequent matching stage.

Let $u$ and $v$ be two currently referenced mapped roots, with selected cuts
$C_u$ and $C_v$. Their input supports and merged support are
\begin{equation}
I_u=\mathrm{supp}(C_u),\qquad
I_v=\mathrm{supp}(C_v),\qquad
I_{uv}=I_u\cup I_v.
\label{eq:score_supports}
\end{equation}
The number of shared inputs is
\begin{equation}
n_{\mathrm{shared}}(u,v)
=
|I_u| + |I_v| - |I_{uv}|.
\label{eq:shared_input_count}
\end{equation}

The primary objective of pairing is to reduce the estimated occupied LUT-site
count.
Thus, the first term in the score is the estimated area saving obtained by
implementing $u$ and $v$ using one dual-output LUT instead of two separate
single-output LUTs:
\begin{equation}
\Delta A(u,v)=A(C_u)+A(C_v)-A_{\mathrm{dual}}(I_{uv})>0.
\label{eq:area_saving}
\end{equation}
where $A(C_u)$ and $A(C_v)$ are the costs of the two individual LUTs, and
$A_{\mathrm{dual}}(I_{uv})$ is the physical cost of the merged dual-output LUT
with input set $I_{uv}$.

In addition to area saving, the score favors pairs with stronger input
sharing. A larger shared input count usually indicates a more natural
fracturable LUT packing opportunity, because the two functions use similar
signals and are less likely to increase local routing pressure. The score also
penalizes large merged supports, since a pair using many distinct inputs
consumes more of the physical LUT input capacity and may reduce architectural
flexibility.

Timing is incorporated as a penalty term. Let
$\mathrm{arr}_{\mathrm{orig}}(u)$ and
$\mathrm{arr}_{\mathrm{orig}}(v)$ denote the arrival times before pairing, and
let $\mathrm{arr}_{\mathrm{pair}}(u)$ and
$\mathrm{arr}_{\mathrm{pair}}(v)$ denote the output-specific arrival times
after pairing. The timing penalty is defined as
\begin{equation}
\begin{aligned}
\Delta D(u,v)
=
\max \Big(&0,\,
\max \big(
\mathrm{arr}_{\mathrm{pair}}(u),
\mathrm{arr}_{\mathrm{pair}}(v)
\big)  \\
&-
\max \big(
\mathrm{arr}_{\mathrm{orig}}(u),
\mathrm{arr}_{\mathrm{orig}}(v)
\big)
\Big).
\end{aligned}
\label{eq:delay_penalty}
\end{equation}
Pairs that violate required times are discarded before scoring:
\begin{equation}
\mathrm{arr}_{\mathrm{pair}}(u)\leq\mathrm{req}(u),\qquad
\mathrm{arr}_{\mathrm{pair}}(v)\leq\mathrm{req}(v).
\label{eq:score_required_times}
\end{equation}

The final score combines area saving, shared-input reward, timing penalty, and
structural complexity penalty:
\begin{equation}
\begin{aligned}
S(u,v)
=
&\ \alpha \Delta A(u,v)
+
\beta n_{\mathrm{shared}}(u,v)
-
\gamma \Delta D(u,v) \\
&-
\eta |I_{uv}|
-
\theta |N_{uv}|.
\end{aligned}
\label{eq:pair_score_full}
\end{equation}
where $N_{uv}$ is the set of internal nodes covered by the merged pair. The
coefficients $\alpha$, $\beta$, $\gamma$, $\eta$, and $\theta$ control the
relative importance of area reduction, input sharing, timing preservation,
input usage, and structural complexity.

In the implementation, area saving is assigned the largest weight because the
main purpose of dual-output pairing is to reduce the estimated occupied
LUT-site count.
The shared-input term then breaks ties among pairs with similar area saving,
while timing and complexity penalties prevent the mapper from selecting pairs
that are structurally large or risky for timing. Legal pairs are sorted in
descending score order. For each LUT, only the best bounded number of incident
edges is retained, $|\mathcal{E}_u| \leq n_{\mathrm{pair}}$, where
$\mathcal{E}_u$ is the retained pair-edge set incident to LUT $u$ and
$n_{\mathrm{pair}}$ corresponds to the implementation parameter
$n_{\mathrm{PairsPerObjMax}}$.

The resulting bounded pair graph is then passed to the matching stage. The
mapper first commits mutual-best pairs and then greedily commits remaining
non-conflicting pairs in descending score order. This score-based filtering
keeps the matching problem small while preserving the most promising
dual-output LUT packing opportunities.

Algorithm~\ref{alg:score_matching} gives the score-based matching procedure.
If $q$ candidates are evaluated and $e$ legal edges are retained, scoring costs
$\mathcal{O}(qK)$, sorting costs $\mathcal{O}(e\log e)$, and the two matching
passes cost $\mathcal{O}(e)$.

\begin{algorithm}[!t]
\caption{Pair Scoring and Matching}
\label{alg:score_matching}
\begin{algorithmic}[1]
\STATE \textbf{Input:} candidate sets $\mathcal{P}_u$, selected cuts,
limit $n_{\mathrm{pair}}$
\STATE \textbf{Output:} committed pair set $\mathcal{M}$
\FOR{each root $u$}
    \STATE $\mathcal{E}_u\leftarrow\emptyset$
    \FOR{each candidate $v\in\mathcal{P}_u$}
        \IF{Algorithm~\ref{alg:legality_check} returns legal for $(u,v)$}
            \STATE Compute $\Delta A(u,v)$, $\Delta D(u,v)$, and $S(u,v)$
            \IF{$\Delta A(u,v)>0$}
                \STATE Insert $(u,v,S(u,v))$ into $\mathcal{E}_u$
            \ENDIF
        \ENDIF
    \ENDFOR
    \STATE Retain the top $n_{\mathrm{pair}}$ edges in $\mathcal{E}_u$
\ENDFOR
\STATE $\mathcal{E}\leftarrow\bigcup_u\mathcal{E}_u$ sorted by descending score
\STATE $\mathcal{M}\leftarrow\emptyset$ and mark all roots unmatched
\FOR{each edge $(u,v)\in\mathcal{E}$}
    \IF{$(u,v)$ is the best edge of both endpoints and both are unmatched}
        \STATE Add $(u,v)$ to $\mathcal{M}$; mark $u$ and $v$ matched
    \ENDIF
\ENDFOR
\FOR{each edge $(u,v)\in\mathcal{E}$ in descending score order}
    \IF{$u$ and $v$ are both unmatched}
        \STATE Add $(u,v)$ to $\mathcal{M}$; mark $u$ and $v$ matched
    \ENDIF
\ENDFOR
\RETURN $\mathcal{M}$
\end{algorithmic}
\end{algorithm}

\section{Experimental Setup}
\label{sec:experimental_setup}

This section outlines the evaluation methodology, including the hardware and software verification platform, the target fracturable LUT architectures, the parameter-scan configurations, and the benchmark suites along with their respective evaluation metrics.

\begin{table}[!t]
\centering
\caption{Representative fracturable-LUT architecture models. Neither model
imposes an additional input-sharing constraint.}
\label{tab:fpga_architectures}

\small
\setlength{\tabcolsep}{8pt}
\renewcommand{\arraystretch}{1.1}

\begin{tabular}{@{}lccccc@{}}
\toprule
Architecture &
$K_{\mathrm{single}}$ &
$K_{\mathrm{out}}$ &
$K_{\mathrm{union}}$ &
$S_{\min}$ &
$S_{\max}$ \\
\midrule
ultrascale+ & 6 & 5 & 5 & 0 & 5 \\
Versal & 6 & 6 & 6 & 0 & 6 \\
\bottomrule
\end{tabular}
\end{table}

\subsection{Experimental Platform}

To evaluate the performance of the proposed dual-output-aware LUT mapping framework, all algorithms are implemented in C++ within the ABC synthesis environment. The experimental evaluations are executed on a 64-bit Windows workstation powered by a 12th Gen Intel(R) Core(TM) i7-12700K processor operating at 3.60~GHz, equipped with 32.0~GB of physical memory.

\subsection{Target FPGA Architectures}

We evaluate two representative architecture models rather than claim an exact
implementation of a particular commercial device.  Both allow six inputs in
single-output mode.  The restricted model limits each dual-output function
and their physical union to five inputs, whereas the Versal-like model permits up
to six inputs for each quantity.  Neither model adds a nontrivial sharing
constraint, so $S_{\min}=0$ and $S_{\max}=K_{\mathrm{out}}$.  Legality follows
Eqs.~\eqref{eq:arch_pair_limits}--\eqref{eq:arch_root_independence}.  The
labels ``UltraScale-like'' and ``Versal-like'' retained inside the
generated plots correspond to the UltraScale-like and Versal-like abstract models,
respectively; they are plot identifiers and should not be interpreted as a
complete model of those commercial devices.

\subsection{Mapping Parameters and Scan Ranges}

The proposed mapper exposes both conventional ABC mapping parameters and dual-output-specific knobs. To systematically characterize their algorithmic impacts, we establish a robust baseline configuration and execute multi-dimensional individual or joint parameter sweeps. The absolute taxonomy, baseline settings, and empirical scan intervals are encapsulated in Table~\ref{tab:mapping_parameters}. During any isolated parameter sweep, all non-scanned parameters are strictly locked at their default baseline values to isolate variables.

\begin{table*}[!t]
\centering
\caption{Taxonomy, baseline settings, and scan ranges of mapping parameters.}
\label{tab:mapping_parameters}
\setlength{\tabcolsep}{16pt} 
\renewcommand{\arraystretch}{1.15}
\begin{tabular}{lllcc}
\toprule
Category & Parameter & Description & Baseline & Scan Range / Set \\
\midrule
\multirow{4}{*}{\shortstack[l]{Conventional\\ABC}} 
& $K$ & LUT size & 6 & --- \\
& $n_F$ & Number of area-flow rounds & 1 & \multirow{2}{*}{$\{0, 1, 2, \dots, 9\}$ (Joint)} \\
& $n_A$ & Number of exact-area rounds & 5 & \\
& $n_{\mathit{cuts}}$ & Max cuts retained per node & 24 & $\{1, 2, 4, 6, 8, 12, 16, 24, 32, 48, 64, 96, 128\}$ \\
\midrule
\multirow{3}{*}{\shortstack[l]{Dual-Output\\Specific}} 
& $n_{\mathit{bucket}}$ & Bucket scan limit & 16 & $\{4, 8, 16, 32, 64\}$ (Joint) \\
& $n_{\mathit{cand}}$ & Candidate limit & 24 & $\{4, 8, 16, 32, 64\}$ (Joint) \\
& $n_{\mathit{pair}}$ & Retained pair limit per object & 16 & $\{2, 4, 8, 16, 32, 64\}$ \\
\bottomrule
\end{tabular}
\end{table*}

\subsection{Benchmarks and Evaluation Metrics}

Our evaluation utilizes standard combinational designs from the EPFL benchmark suite. The reference baseline is generated by running the vanilla, single-output ABC mapping flow to output a standard BLIF representation. The proposed dual-output-aware mapper is then driven by the same source netlists under varying parameter configurations. 

For an integer site-count evaluation, let $N_{\mathrm{single}}$ be the number
of unpaired LUT roots and $N_{\mathrm{dual}}$ the number of committed root
pairs.  The occupied physical-site count and dual-output-site utilization are
\begin{equation}
N_{\mathrm{phys}}=N_{\mathrm{single}}+N_{\mathrm{dual}},\qquad
R_{\mathrm{dual-site}}=\frac{N_{\mathrm{dual}}}{N_{\mathrm{phys}}}\times100\%.
\label{eq:site_metrics}
\end{equation}
The source result image reports some LUT-area values at half-site precision~\cite{Lu_2025_MOLUT}.
To provide an integer comparison table, those values are rounded to the
nearest integer in Tables~\ref{tab:benchmark_versal} and
\ref{tab:benchmark_restricted}; percentages and depths are copied unchanged.
The resulting $N_{\mathrm{prop}}$ column is therefore an integer estimate
derived from the reported data.

For the currently available results, the conventional single-output ABC
solution is the area baseline. The reported normalized area reduction,
dual-output percentage, and structural depth gain are
\begin{equation}
\begin{aligned}
G_{\mathrm{area}}&=\frac{N_{\mathrm{ABC}}-N_{\mathrm{prop}}}{N_{\mathrm{ABC}}}\times100\%,\\
R_{\mathrm{dual}}&=\frac{N_{\mathrm{dual}}}{N_{\mathrm{prop}}}\times100\%,\\
G_{\mathrm{depth}}&=\frac{D_{\mathrm{baseline}}-D_{\mathrm{prop}}}{D_{\mathrm{baseline}}}\times100\%.
\end{aligned}
\label{eq:evaluation_metrics}
\end{equation}
Here, $N_{\mathrm{ABC}}$ is the ABC LUT count, $N_{\mathrm{prop}}$ is the
rounded estimated occupied-LUT count reported in the detailed tables, and
$N_{\mathrm{dual}}$ is the reported number of packed pairs. Execution time is
reported in seconds.
The present experiments evaluate structural mapping depth; they do not report
post-place-and-route critical delay, slack, or a calibrated commercial-device
STA result.

\section{Experimental Result Analysis}
\label{sec:experimental_results}

\subsection{Benchmark-Level Comparison}
\label{subsec:benchmark_comparison}

Tables~\ref{tab:benchmark_versal} and \ref{tab:benchmark_restricted} report
the complete per-benchmark results for both target architectures.  The reference columns reproduce the proposed
method's simulation results obtained with the original benchmark setup~\cite{Lu_2025_MOLUT},
whereas the current-local-results columns report the present implementation
under the experimental configuration described above.  For Versal-like
mapping, the current implementation achieves an average LUT-area saving of
34.96\%, compared with 33.91\% in the reference data, while preserving a
similar dual-output percentage (58.68\% versus 59.10\%).  The corresponding
average depth improvement is 3.38\%, and the mean runtime is 3.801~s.  For the
more restrictive UltraScale-like architecture, the average LUT-area
saving increases from 22.40\% to 23.39\%, the dual-output percentage changes
from 35.03\% to 35.41\%, and the current implementation improves depth by
5.13\% on average with a mean runtime of 3.571~s.

The benchmark-level data also reveal substantial circuit-dependent
variation.  Designs such as \texttt{div}, \texttt{sin}, and \texttt{voter}
show clear depth or area improvements, whereas \texttt{hyp} remains a
difficult case and uses more LUT area than the reference result under both
architectures.  This variation motivates reporting the complete data rather
than relying only on suite averages, and it also indicates that future
scoring refinements should account for benchmark structures with limited
high-quality pairing alternatives.

\begin{table*}[!t]
\centering
\caption{Benchmark-level results for versal architecture.}
\label{tab:benchmark_versal}
\scriptsize
\setlength{\tabcolsep}{9pt}
\renewcommand{\arraystretch}{1.05}
\begin{tabular}{lrrrrr|rrrrrr}
\toprule
& \multicolumn{5}{c|}{Reference results~\cite{Lu_2025_MOLUT}}
& \multicolumn{6}{c}{Current results} \\
Benchmark
& LUTs & Depth & Dual (\%) & Save (\%) & Time (s)
& LUTs & Depth & $G_D$ (\%) & Dual (\%) & Save (\%) & Time (s) \\
\midrule
adder      & 152   & 51   & 70.39 & 40.16 & 0.07
           & 154   & 51   & 0.00  & 82.41 & 40.27 & 0.04 \\

arbiter    & 2467  & 18   & 10.34 & 9.37  & 2.51
           & 2465  & 18   & 0.00  & 10.43 & 9.44  & 0.54 \\

bar        & 448   & 4    & 14.29 & 12.50 & 0.27
           & 448   & 4    & 0.00  & 14.29 & 12.50 & 0.11 \\

cavlc      & 80    & 4    & 45.00 & 31.03 & 0.06
           & 74    & 4    & 0.00  & 56.76 & 36.21 & 0.04 \\

ctrl       & 16    & 2    & 81.25 & 44.83 & 0.01
           & 16    & 2    & 0.00  & 81.25 & 44.83 & 0.01 \\

dec        & 144   & 2    & 99.31 & 49.83 & 0.15
           & 151   & 2    & 0.00  & 90.07 & 47.39 & 0.03 \\

div        & 14155 & 936  & 83.64 & 35.99 & 274.31
           & 12160 & 864  & 7.69  & 84.16 & 45.36 & 15.06 \\

hyp        & 24983 & 4214 & 80.07 & 43.87 & 762.75
           & 27689 & 4194 & 0.47  & 62.24 & 37.77 & 36.57 \\

i2c        & 259   & 4    & 33.98 & 26.63 & 0.13
           & 248   & 4    & 0.00  & 40.20 & 28.67 & 0.06 \\

int2float  & 34    & 3    & 47.06 & 33.33 & 0.01
           & 36    & 3    & 0.00  & 44.44 & 30.77 & 0.01 \\

log2       & 5330  & 74   & 49.25 & 34.13 & 32.94
           & 5409  & 71   & 4.05  & 49.75 & 32.15 & 5.30 \\

max        & 513   & 45   & 52.05 & 33.29 & 0.43
           & 507   & 44   & 2.22  & 51.48 & 33.98 & 0.15 \\

mem\_ctrl  & 8482  & 27   & 45.59 & 29.81 & 63.53
           & 8391  & 25   & 7.41  & 42.97 & 30.12 & 5.49 \\

multiplier & 3790  & 58   & 56.07 & 36.06 & 22.53
           & 3763  & 53   & 8.62  & 58.70 & 35.77 & 4.38 \\

priority   & 123   & 31   & 74.80 & 41.43 & 0.07
           & 119   & 31   & 0.00  & 75.53 & 44.37 & 0.04 \\

router     & 78    & 7    & 30.77 & 12.36 & 0.01
           & 72    & 7    & 0.00  & 37.06 & 26.29 & 0.01 \\

sin        & 957   & 40   & 54.13 & 34.63 & 2.14
           & 928   & 36   & 10.00 & 59.57 & 36.73 & 0.79 \\

sqrt       & 4236  & 1025 & 72.85 & 25.83 & 13.00
           & 3840  & 1024 & 0.10  & 79.06 & 39.81 & 2.60 \\

square     & 2041  & 51   & 96.28 & 48.95 & 15.63
           & 2352  & 50   & 1.96  & 68.79 & 40.95 & 3.42 \\

voter      & 1232  & 20   & 84.82 & 54.29 & 4.63
           & 1221  & 15   & 25.00 & 84.43 & 45.76 & 1.38 \\

\midrule
Average    & -- & -- & 59.10 & 33.91 & 59.76
           & -- & -- & 3.38 & 58.68 & 34.96 & 3.80 \\
\bottomrule
\end{tabular}
\par\vspace*{2pt}
\begin{minipage}{0.95\textwidth}
\footnotesize
\textit{Note:} $G_D$ denotes the depth reduction achieved by the current method relative to the reference method, calculated as
$G_D=(D_{\mathrm{ref}}-D_{\mathrm{current}})/D_{\mathrm{ref}}\times 100\%$.
Save denotes the reduction in the LUT-area metric relative to the corresponding single-output LUT mapping baseline. 
\end{minipage}
\end{table*}

\begin{table*}[!t]
\centering
\caption{Benchmark-level results for ultrascale+ architecture.}
\label{tab:benchmark_restricted}
\scriptsize
\setlength{\tabcolsep}{9pt}
\renewcommand{\arraystretch}{1.05}
\begin{tabular}{lrrrrr|rrrrrr}
\toprule
& \multicolumn{5}{c|}{Reference results~\cite{Lu_2025_MOLUT}}
& \multicolumn{6}{c}{Current results} \\
Benchmark
& LUTs & Depth & Dual (\%) & Save (\%) & Time (s)
& LUTs & Depth & $G_D$ (\%) & Dual (\%) & Save (\%) & Time (s) \\
\midrule
adder      & 183   & 51   & 40.44 & 27.95 & 0.08
           & 181   & 51   & 0.00  & 69.06 & 29.57 & 0.04 \\

arbiter    & 2468  & 18   & 10.29 & 9.33  & 2.00
           & 2467  & 18   & 0.00  & 10.36 & 9.39  & 0.41 \\

bar        & 448   & 4    & 14.29 & 12.50 & 0.23
           & 448   & 4    & 0.00  & 14.29 & 12.50 & 0.09 \\

cavlc      & 108   & 4    & 7.41  & 6.90  & 0.05
           & 102   & 4    & 0.00  & 13.73 & 12.07 & 0.03 \\

ctrl       & 18    & 2    & 61.11 & 37.93 & 0.01
           & 18    & 2    & 0.00  & 61.11 & 37.93 & 0.01 \\

dec        & 144   & 2    & 99.31 & 49.83 & 0.15
           & 150   & 2    & 0.00  & 91.33 & 47.74 & 0.03 \\

div        & 16289 & 943  & 57.97 & 26.34 & 244.69
           & 13479 & 864  & 8.38  & 65.84 & 39.44 & 15.37 \\

hyp        & 25007 & 4214 & 79.90 & 43.81 & 753.49
           & 33947 & 4194 & 0.47  & 31.46 & 23.70 & 34.32 \\

i2c        & 283   & 5    & 25.09 & 19.83 & 0.11
           & 283   & 4    & 20.00 & 22.83 & 18.59 & 0.05 \\

int2float  & 44    & 3    & 15.91 & 13.73 & 0.01
           & 42    & 3    & 0.00  & 23.81 & 19.23 & 0.01 \\

log2       & 6412  & 77   & 29.43 & 20.76 & 32.58
           & 6249  & 71   & 7.79  & 28.32 & 21.61 & 4.98 \\

max        & 571   & 45   & 36.60 & 25.75 & 0.39
           & 556   & 44   & 2.22  & 38.25 & 27.67 & 0.13 \\

mem\_ctrl  & 9773  & 30   & 27.79 & 19.12 & 49.45
           & 9326  & 25   & 16.67 & 28.72 & 22.34 & 4.06 \\

multiplier & 4398  & 56   & 36.02 & 25.80 & 22.85
           & 4365  & 53   & 5.36  & 36.99 & 25.50 & 4.22 \\

priority   & 184   & 31   & 15.76 & 12.38 & 0.07
           & 182   & 31   & 0.00  & 16.25 & 14.79 & 0.04 \\

router     & 76    & 7    & 17.11 & 14.61 & 0.01
           & 81    & 7    & 0.00  & 20.99 & 16.49 & 0.01 \\

sin        & 1198  & 41   & 24.46 & 18.17 & 2.14
           & 1149  & 36   & 12.20 & 28.11 & 21.62 & 0.75 \\

sqrt       & 4913  & 1051 & 28.25 & 13.97 & 11.05
           & 5201  & 1024 & 2.57  & 38.26 & 18.49 & 2.10 \\

square     & 2717  & 51   & 48.25 & 32.04 & 16.51
           & 2719  & 50   & 1.96  & 47.58 & 31.73 & 3.43 \\

voter      & 2228  & 20   & 25.22 & 17.33 & 4.77
           & 1860  & 15   & 25.00 & 21.00 & 17.36 & 1.35 \\

\midrule
Average    & -- & -- & 35.03 & 22.40 & 57.03
           & -- & -- & 5.13 & 35.41 & 23.39 & 3.57 \\
\bottomrule
\end{tabular}
\par\vspace*{2pt}
\begin{minipage}{0.95\textwidth}
\footnotesize
\textit{Note:} $G_D$ denotes the depth reduction achieved by the current method relative to the reference method, calculated as
$G_D=(D_{\mathrm{ref}}-D_{\mathrm{current}})/D_{\mathrm{ref}}\times 100\%$.
Save denotes the reduction in the LUT-area metric relative to the corresponding single-output LUT mapping baseline. 
\end{minipage}
\end{table*}

\subsection{Dual-Output Parameter Analysis}
\label{subsec:dual_output_param_analysis}

This subsection evaluates the joint and individual impacts of key tuning parameters on the proposed dual-output-aware LUT mapping flow, exploring how candidate generation and pair retention limits govern the final mapping quality.

\begin{figure}[!t]
    \centering
    \includegraphics[width=1\linewidth]{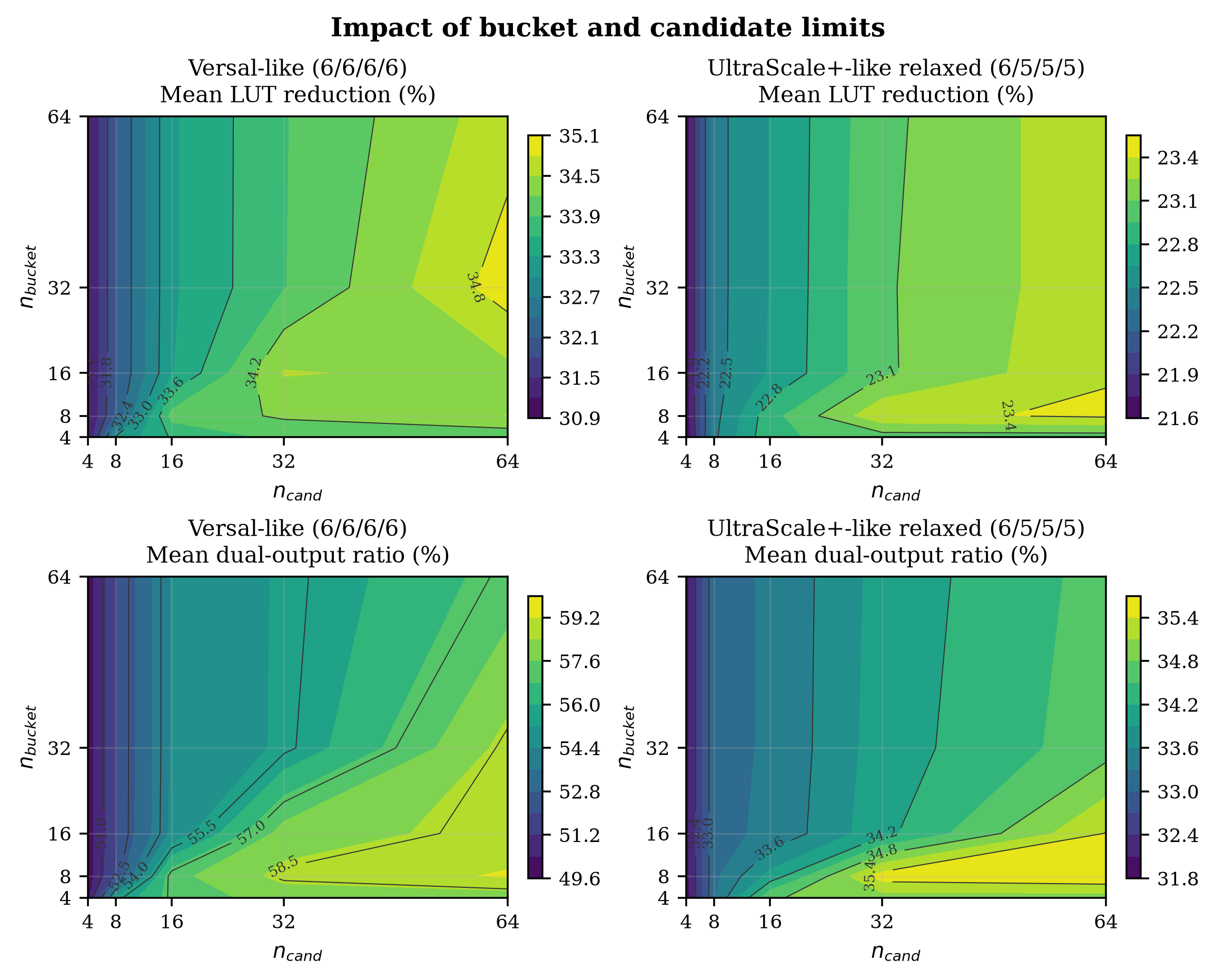}
    \caption{Impact of $n_{\mathrm{bucket}}$ and $n_{\mathrm{cand}}$ on
    dual-output-aware mapping for both architectures.  The top row reports
    mean LUT reduction and the bottom row reports mean dual-output LUT ratio;
    the left and right columns correspond to Versal-like and UltraScale-like
    architectures, respectively.}
    \label{fig:bucket_candidate}
\end{figure}

\subsubsection{Impact of Bucket and Candidate Limits}
\label{subsubsec:bucket_candidate_results}

This section evaluates the joint impact of the shared-input bucket scan limit $n_{\mathrm{bucket}}$ and candidate limit $n_{\mathrm{cand}}$ on mapping quality, as these two parameters directly dictate the candidate-generation stage. Fig.~\ref{fig:bucket_candidate} illustrates the variations in mean LUT reduction and dual-output LUT ratio across the benchmark suite. Conceptually, $n_{\mathrm{bucket}}$ controls the local search depth within each shared-input bucket, whereas $n_{\mathrm{cand}}$ bounds the global candidate pool size fed into subsequent legality checking and scoring. 

For both architectures, the most pronounced improvement occurs as
$n_{\mathrm{cand}}$ increases from 4 to 16.  Beyond 32, the contours become
nearly vertical and the incremental gain is small.  At the largest tested
candidate budgets, the Versal-like architecture reaches approximately
34--35\% mean LUT reduction and 58--59\% mean dual-output ratio, whereas the
more restrictive UltraScale-like architecture reaches approximately
23--23.5\% and 34--35.5\%, respectively.  The consistent separation between
the two architectures quantifies the benefit of the larger legal dual-output
support offered by the Versal-like mode.

The bucket limit has a weaker, non-monotonic effect.  With a sufficiently
large candidate budget, moderate values around $n_{\mathrm{bucket}}=8$--16
often yield the highest dual-output ratio, while larger bucket scans provide
little additional area benefit and can slightly dilute the retained candidate
set.  Thus, increasing the global candidate budget is useful only up to the
point at which the local shared-input neighborhoods are adequately covered.

Importantly, the results demonstrate that achieving superior area efficiency relies on the synergistic optimization of both parameters under a reasonable configuration. Furthermore, the system's global area savings exhibit a strong positive correlation with the dual-output LUT percentage. This robust positive correlation indicates that the proposed mapping framework successfully translates structural dual-output opportunities into tangible hardware reduction, rather than introducing redundant or low-quality logic duplication. This further implies that our bounded candidate search space is structurally high-yielding, confirming that an intelligent, localized clustering metric can effectively substitute for a computationally prohibitive global search while maintaining strict mapping quality.

\begin{figure}[!t]
    \centering
    \includegraphics[width=0.9\linewidth]{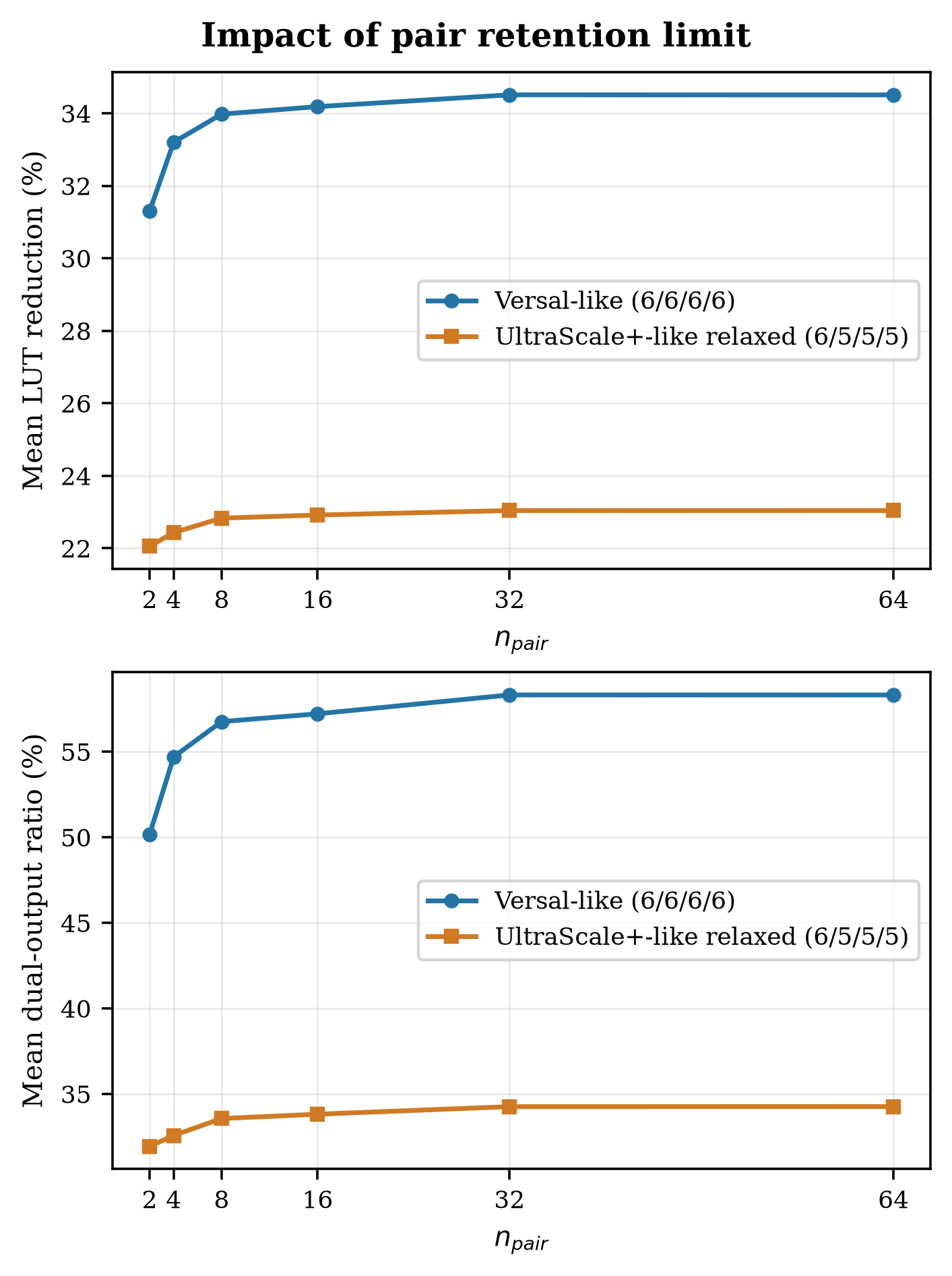}
    \caption{Impact of $n_{\mathrm{pair}}$ on both target architectures:
    (a) mean LUT reduction and (b) mean dual-output LUT ratio.}
    \label{fig:pair_limit}
\end{figure}

\subsubsection{Impact of Pair Retention Limit}
\label{subsubsec:pair_limit_results}

This section evaluates the impact of the retained pair limit per object $n_{\mathrm{pair}}$ on mapping quality. This parameter controls the maximum number of high-scoring legal pair edges preserved for each LUT before the matching stage. Fig.~\ref{fig:pair_limit} reports the mean LUT reduction and dual-output LUT ratio under different $n_{\mathrm{pair}}$ values.

The results show that mapping quality is most sensitive to
$n_{\mathrm{pair}}$ when the retained edge budget is small.  Increasing
$n_{\mathrm{pair}}$ from 2 to 8 raises the Versal-like mean LUT reduction from
about 31.3\% to 34.0\% and its dual-output ratio from 50.0\% to about 56.7\%.
Over the same range, UltraScale-like improves from about 22.1\% to
22.8\% LUT reduction and from about 32.0\% to 33.6\% dual-output ratio.  Thus,
early pruning is more consequential for the less restrictive architecture,
whose larger legal-pair space requires more alternatives during conflict
resolution.

After $n_{\mathrm{pair}}$ reaches 16--32, both architectures saturate.  At
$n_{\mathrm{pair}}=32$, the Versal-like results are about 34.5\% LUT reduction
and 58.2\% dual-output ratio, while UltraScale-like reaches about
23.0\% and 34.3\%, respectively.  Increasing the limit to 64 produces no
visible additional gain, showing that a compact retained graph contains the
useful matching alternatives and that $n_{\mathrm{pair}}=16$ or 32 is a
practical quality--complexity compromise.

\subsection{ABC Original Parameter Analysis}
\label{subsec:abc_original_parameter_analysis}

\subsubsection{Impact of Maximum Cut Count}
\label{subsubsec:abc_original_cut_count}

This section analyzes the impact of the original ABC parameter
$n_{\mathrm{cuts}}$. This parameter determines the maximum number of priority cuts retained for each
node during cut-based LUT mapping. A larger $n_{\mathrm{cuts}}$ provides a
larger cut search space, but also increases mapping runtime.

\begin{figure}[!t]
    \centering
    \includegraphics[width=1\linewidth]{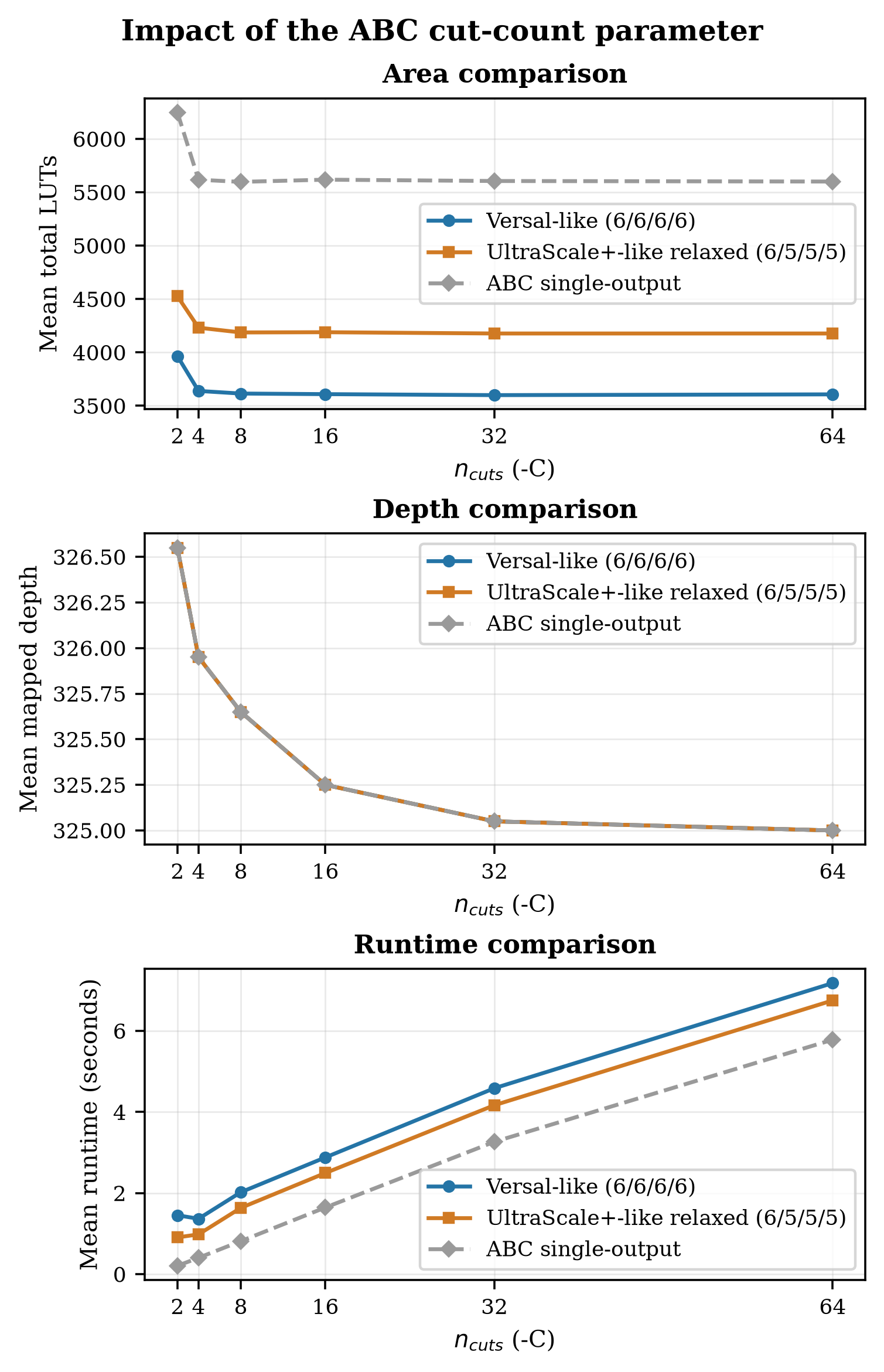}
    \caption{Impact of the original ABC cut-count parameter
    $n_{\mathrm{cuts}}$ on ABC single-output, Versal-like, and
    UltraScale-like mapping: (a) mean total LUT count, (b) mean
    mapped depth, and (c) mean runtime.}
    \label{fig:abc_cut_analysis}
\end{figure}

Fig.~\ref{fig:abc_cut_analysis} compares the proposed dual-output mapper with
the original ABC single-output mapper under different $n_{\mathrm{cuts}}$
settings. As shown in Fig.~\ref{fig:abc_cut_analysis}(a), increasing
$n_{\mathrm{cuts}}$ improves the proposed mapper when the cut budget is small.
The mean total LUT count decreases rapidly from $n_{\mathrm{cuts}}=2$ to
$n_{\mathrm{cuts}}=8$, and then gradually saturates. This indicates that the
dual-output mapper benefits from a sufficiently rich cut set, because more
cuts provide more legal and high-quality pairing opportunities. However, after
$n_{\mathrm{cuts}}\geq 16$, the area improvement becomes marginal.

The original ABC result is much less sensitive to $n_{\mathrm{cuts}}$ in the
same range.  After the initial drop, it remains near 5600 mean LUTs, compared
with approximately 3600 for Versal-like and 4180 for UltraScale-like
at moderate and large cut budgets.  The persistent gap between both proposed
configurations and ABC shows that the area advantage mainly comes from
dual-output-aware mapping and feedback rather than only from retaining more
priority cuts.  The additional gap between the two proposed curves is the
cost of the tighter UltraScale-like legality constraints.

Fig.~\ref{fig:abc_cut_analysis}(b) shows that both flows have almost identical
mapped depth across all $n_{\mathrm{cuts}}$ settings. The depth decreases
slightly as more cuts are retained and then converges after
$n_{\mathrm{cuts}}\geq 16$. This confirms that the proposed dual-output
optimization preserves the structural mapping depth of the original ABC flow
while reducing the reported LUT-area metric.

Fig.~\ref{fig:abc_cut_analysis}(c) reports the runtime comparison. Runtime
increases with $n_{\mathrm{cuts}}$ for both methods, but the proposed mapper
has a higher slope because each mapping round additionally performs
dual-output candidate generation, legality checking, scoring, and matching.
Therefore, $n_{\mathrm{cuts}}$ controls an important quality-runtime trade-off.
A moderate value, such as $n_{\mathrm{cuts}}=16$ or $32$, provides most of the
area benefit while avoiding the high runtime cost observed at larger cut
budgets.

\section{Limitations}
\label{sec:limitations}

The present evaluation has four important limitations.  First, shared-input
indexing is a heuristic primary candidate source.  When $S_{\min}=0$, legal
disjoint-support pairs may be omitted, and the bounded matching heuristic does
not guarantee a maximum-cardinality or maximum-weight pair set.  Exhaustive
candidate recall and optimal-matching comparisons remain necessary on
tractable benchmarks to quantify this loss.

Second, the available experiments compare the proposed flow with vanilla ABC
but do not yet include an ABC-plus-post-packing baseline using identical
legality, scoring, and matching.  Consequently, the current data demonstrate
the combined benefit of dual-output packing and iterative mapping, but do not
isolate the incremental contribution of feedback.

Third, timing is evaluated using structural mapping depth and output-specific
logical supports.  No calibrated pin-dependent LUT delays, post-routing
critical-path delay, worst negative slack, or comparison with union-support
timing is currently available.  The results therefore support preservation of
mapping depth under the evaluated model, not a claim of commercial-device
timing closure.

Finally, the integer LUT counts in Tables~\ref{tab:benchmark_versal} and
\ref{tab:benchmark_restricted} are rounded from the available source data.
Directly regenerated raw single- and dual-site counts would avoid this
rounding step.  Repeated-run statistics, source revision, compiler options,
benchmark preprocessing commands, and checksums also remain to be documented.

\section{Conclusion}
\label{sec:conclusion}

This paper presented an iterative dual-output-aware LUT mapping framework for
fracturable FPGA architecture models.  The method alternates between
single-output cut selection and bounded heuristic pair matching and feeds pair
compatibility into subsequent area-recovery rounds.  A unified legality model
separates per-output support limits, physical input-union limits, optional
sharing constraints, root independence, and output-specific logical timing
supports.

On the currently available EPFL results, the reported LUT-area metric is
reduced by 34.96\% and 23.39\% on average relative to vanilla ABC for the
Versal-like and UltraScale-like models, respectively.  Average structural-depth gains
are 3.38\% and 5.13\%.  These results show that bounded iterative pairing can
expose useful resource-sharing opportunities, but they do not yet isolate
feedback from post-packing or establish routed timing performance.  Future
work will evaluate union-aware candidate indexing, exhaustive and
maximum-weight matching on tractable designs, architecture-calibrated delay
models, and placement- and routing-aware dual-output mapping.

\balance
\bibliographystyle{IEEEtran}
\bibliography{citations_revised}

\end{document}